\documentclass[twoside,twocolumn,american,prb,superscriptaddress,showpacs]{revtex4}
\usepackage[T1]{fontenc}
\usepackage[latin9]{inputenc}
\usepackage{amsmath}
\usepackage{graphicx}
\usepackage{amssymb}

\makeatletter

\newcommand{\noun}[1]{\textsc{#1}}
\providecommand{\tabularnewline}{\\}

\@ifundefined{textcolor}{}
{%
 \definecolor{BLACK}{gray}{0}
 \definecolor{WHITE}{gray}{1}
 \definecolor{RED}{rgb}{1,0,0}
 \definecolor{GREEN}{rgb}{0,1,0}
 \definecolor{BLUE}{rgb}{0,0,1}
 \definecolor{CYAN}{cmyk}{1,0,0,0}
 \definecolor{MAGENTA}{cmyk}{0,1,0,0}
 \definecolor{YELLOW}{cmyk}{0,0,1,0}
 }

\usepackage{amsthm}\@ifundefined{definecolor}
 {\usepackage{color}}{}
\usepackage{soul}\usepackage{float}\usepackage{babel}

\makeatletter

\@ifundefined{textcolor}{}
{%
 \definecolor{BLACK}{gray}{0}
 \definecolor{WHITE}{gray}{1}
 \definecolor{RED}{rgb}{1,0,0}
 \definecolor{GREEN}{rgb}{0,1,0}
 \definecolor{BLUE}{rgb}{0,0,1}
 \definecolor{CYAN}{cmyk}{1,0,0,0}
 \definecolor{MAGENTA}{cmyk}{0,1,0,0}
 \definecolor{YELLOW}{cmyk}{0,0,1,0}
 }

\makeatother

\makeatother

\usepackage{babel}

\begin{document}

\title{Conduction mechanisms in biphenyl-dithiol single-molecule junctions}

\author{M.~B\"urkle}

\affiliation{Institute of Theoretical Solid State Physics, Karlsruhe Institute
of Technology, 76131 Karlsruhe, Germany}

\affiliation{Center for Functional Nanostructures, Karlsruhe Institute of Technology,
76131 Karlsruhe, Germany}

\author{J. ~K.~Viljas}

\affiliation{Low Temperature Laboratory, Aalto University, P.O. Box 15100, 00076
AALTO, Finland}

\author{A.~Mishchenko}

\affiliation{Department of Chemistry and Biochemistry, University of Bern, 3012
Bern, Switzerland}

\author{D.~Vonlanthen}

\affiliation{Department of Chemistry, University of Basel, 4003 Basel, Switzerland}

\author{G.~Sch\"on}

\affiliation{Institute of Theoretical Solid State Physics, Karlsruhe Institute
of Technology, 76131 Karlsruhe, Germany}

\affiliation{Center for Functional Nanostructures, Karlsruhe Institute of Technology,
76131 Karlsruhe, Germany}

\affiliation{Institute of Nanotechnology, Karlsruhe Institute of Technology, 76344
Eggenstein-Leopoldshafen, Germany}

\author{M.~Mayor}

\affiliation{Department of Chemistry, University of Basel, 4003 Basel, Switzerland}

\affiliation{Institute of Nanotechnology, Karlsruhe Institute of Technology, 76344
Eggenstein-Leopoldshafen, Germany}

\affiliation{Center for Functional Nanostructures, Karlsruhe Institute of Technology,
76131 Karlsruhe, Germany}

\email{marcel.mayor@unibas.ch}

\author{T.~Wandlowski}

\affiliation{Department of Chemistry and Biochemistry, University of Bern, 3012
Bern, Switzerland}

\email{thomas.wandlowski@dcb.unibe.ch}

\author{F.~Pauly}

\affiliation{Institute of Theoretical Solid State Physics, Karlsruhe Institute
of Technology, 76131 Karlsruhe, Germany}

\affiliation{Center for Functional Nanostructures, Karlsruhe Institute of Technology,
76131 Karlsruhe, Germany}

\email{fabian.pauly@kit.edu}
\begin{abstract}
Based on density-functional theory calculations, we report a detailed
study of the single-molecule charge-transport properties for a series
of recently synthesized biphenyl-dithiol molecules {[}D. Vonlanthen
\emph{et al.}, Angew. Chem., Int. Ed. \textbf{48}, 8886 (2009); A.
Mishchenko \emph{et al.}, Nano Lett. \textbf{10}, 156 (2010){]}. The
torsion angle $\varphi$ between the two phenyl rings, and hence the
degree of $\pi$ conjugation, is controlled by alkyl chains and methyl
side groups. We consider three different coordination geometries,
namely top-top, bridge-bridge, and hollow-hollow with the terminal
sulfur atoms bound to one, two, and three gold surface atoms, respectively.
Our calculations show that different coordination geometries give
rise to conductances which vary by one order of magnitude for the
same molecule. Irrespective of the coordination geometries, the charge
transport calculations predict a $\cos^{2}\varphi$ dependence of
the conductance, which is confirmed by our experimental measurements.
We observe that the calculated transmission through biphenyl dithiols
is typically dominated by a single transmission eigenchannel formed
from $\pi$ electrons. Only for a single molecule with a completely
broken conjugation we find a perfect channel degeneracy for the hollow-hollow-type
contact in our theory.
\end{abstract}


\maketitle

\section{Introduction}

After the first realizations of single- or few-molecule contacts,\cite{Reed1997,Reichert2002,Xu2003}
a major theme of research represents the controlled fabrication of
molecular junctions with desired properties. As an example, several
groups demonstrated recently that the conductance of junctions containing
biphenyl derivatives can be controlled by the torsion angle $\varphi$
between the two phenyl rings.\cite{Venkataraman2006,Mishchenko2010,Mishchenko2011}
However, not only the investigation of the conductance of more complex
molecules has become feasible,\cite{Wu2008,Lafferentz2009,Molen2009}
but also the study of additional aspects such as the signature of
molecular vibrations in the electric current,\cite{Smit2002} current-induced
heating,\cite{Ioffe2008,Ward2011} and the distinction of electron
or hole conduction by measurement of the thermopower.\cite{Reddy2007}
Furthermore, information on individual conduction channels can be
obtained by use of superconducting electrodes\cite{Scheer1998} or
shot-noise measurements.\cite{Kiguchi2008} All these advances allow
for a better characterization of the single-molecule charge transport
when compared with theory.%
\begin{figure}[b]
\begin{centering}
\includegraphics[width=0.9\linewidth]{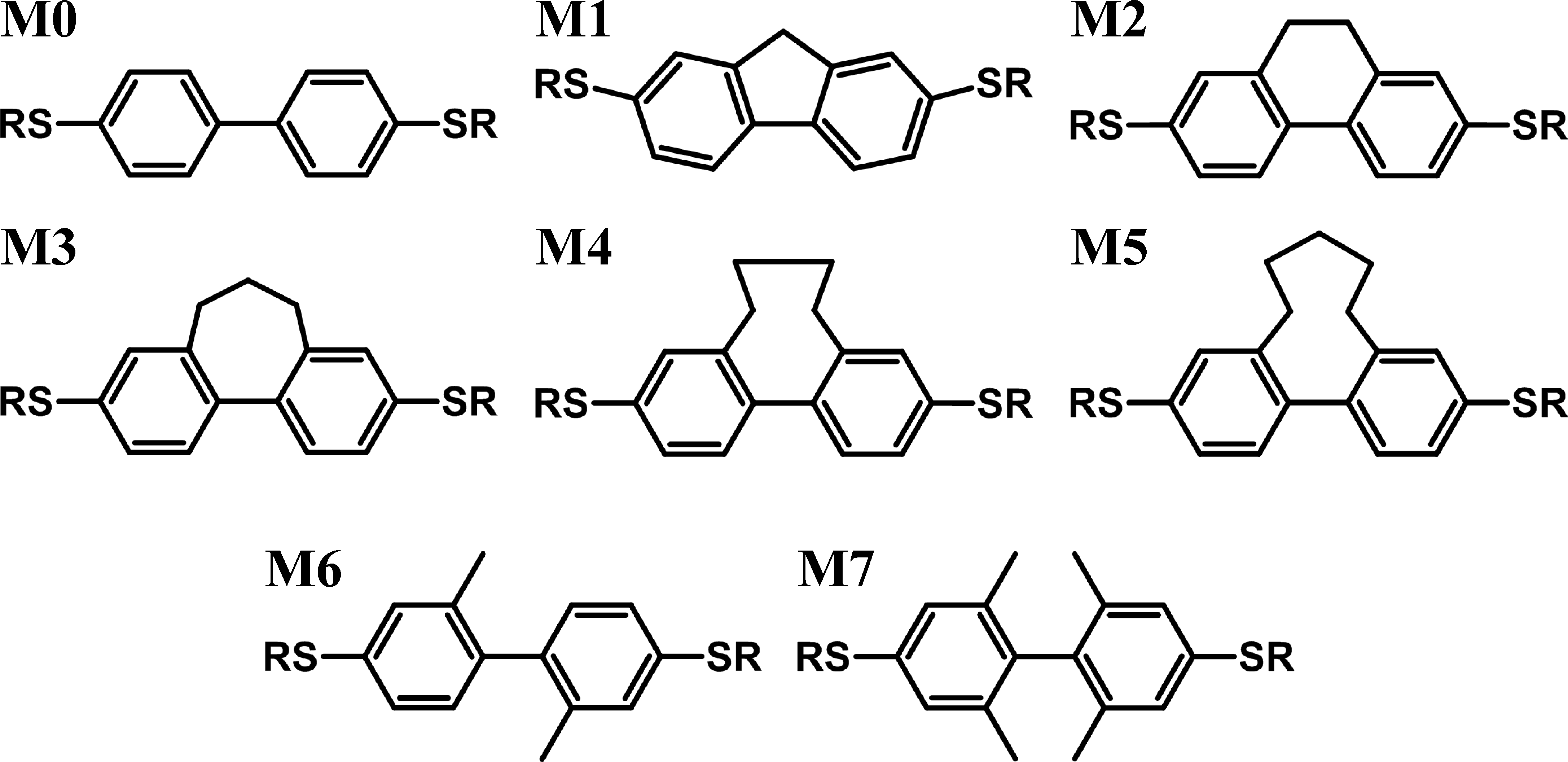} 
\par\end{centering}

\caption{\label{fig:chemstruct}Chemical structure of the investigated molecules.
{}``S'' is the sulfur atom, and {}``R'' represents the acetyl
group for the synthesized form of the molecules, a hydrogen atom after
the in-situ deprotection, or the Au electrode for the transport measurements.}
\end{figure}

Despite experimental and theoretical achievements, measurement and
modeling of electron transport in molecular junctions are still challenging
tasks. This is mainly due to the observed variability in junction
conductances and the corresponding statistical nature of the experiments.\cite{Xu2003,Venkataraman2006a,Li2008}
In this regard, calculations based on the approximate density functional
theory (DFT) can be helpful to obtain a better understanding of the
charge transport mechanisms involved and to interpret trends in the
experimental data based on computed structure-transport relationships.
In agreement with the experimental observations, they show, in particular,
that the electric conduction strongly depends on the molecular conformation\cite{Samanta1996,Venkataraman2006,Pauly2008b,Pauly2008a,Kondo2008,Finch2008,Solomon2009,Mishchenko2010,Mishchenko2011}
and the precise geometry in the single-molecule junctions.\cite{Xue2003,Tomfohr2004,Basch2005,Li2007,Quek2009}

In our recent studies,\cite{Vonlanthen2009,Mishchenko2010} we have
explored the conduction properties of biphenyl-dithiol (BPDT) molecules
bound to Au electrodes. For these molecules, named here M0-M7 and
displayed in Fig.~\ref{fig:chemstruct}, the molecular conjugation
is gradually varied by the use of alkyl chains and methyl side groups.
In this follow-up paper, we present a more detailed theoretical analysis
of their transport properties based on DFT calculations. We study
an extended, systematic set of contact geometries and place special
emphasis on transport for perpendicular ring orientations. The conduction
mechanisms are revealed by means of a tight-binding model (TBM), the
more frequently used Lorentz model (LM), and the eigenchannel decomposition
of the conductance. Our calculations suggest that the coordination
site ({}``top'', {}``bridge'', or {}``hollow'') of the anchoring
sulfur atom at the Au surface plays a decisive role in conduction
through the molecular junction.

The paper is organized as follows. In Sec.~\ref{sec:Experiments},
we briefly summarize key experimental findings.\cite{Vonlanthen2009,Mishchenko2010}
Technical aspects of the DFT and transport calculations are discussed
in Sec.~\ref{sec:Methods}. Studies of the dependence of junction
conductance on the torsion angle and the substrate-adsorbate coordination
geometry are presented and analyzed in Sec.~\ref{sec:Results}. The
main text ends with a summary and conclusions in Sec.~\ref{sec:Conclusions}.
The appendices contain details on the methods used to analyze the
DFT results in Sec.~\ref{sec:Results}. Thus, in Appendix \ref{sec:Teigchans}
we explain our scheme to construct transmission eigenchannel wavefunctions
without the need to resort to L\"owdin transformations when nonorthogonal,
local basis sets are used in the DFT calculations. Finally, Appendix
\ref{sec:Rel-TBM-LM} discusses the relation between the TBM and the
LM.

\section{Experiments\label{sec:Experiments}}

We synthesized the BPDT molecules M0-M7 of Fig.~\ref{fig:chemstruct}
with acetyl-protected terminal thiol groups, i.e.~R=COCH$_{3}$.\cite{Shaporenko2006,Vonlanthen2009,Vonlanthen2010}
The torsion angle $\varphi$ between the phenyl rings is gradually
varied by introduction of alkyl side chains of variable length or
methyl groups.\cite{Vonlanthen2009,Mishchenko2010} This leaves the
length of the molecule unchanged. The torsion angles were determined
by an X-ray structure analysis of single crystals formed from each
of these compounds, except for M0.

This systematic set of molecules exhibits several remarkable features
in single-molecule transport measurements, which we examine further
below. First, stable junctions can be formed with gold leads by the
terminal sulfur atoms after deprotection. Second, the conformation
of the biphenyls is efficiently locked by the alkyl chains and the
steric hindrance of the methyl side groups. Variations of torsion
angles are hence expected to be low.\cite{Pauly2008b,Malen2009a,Rotzler2011}
Third, strongly electron-donating or electron-withdrawing side groups
are avoided, which have been demonstrated to influence noticeably
the single-molecule transport.\cite{Venkataraman2007}

\begin{figure}
\centering{}\includegraphics[width=1\columnwidth]{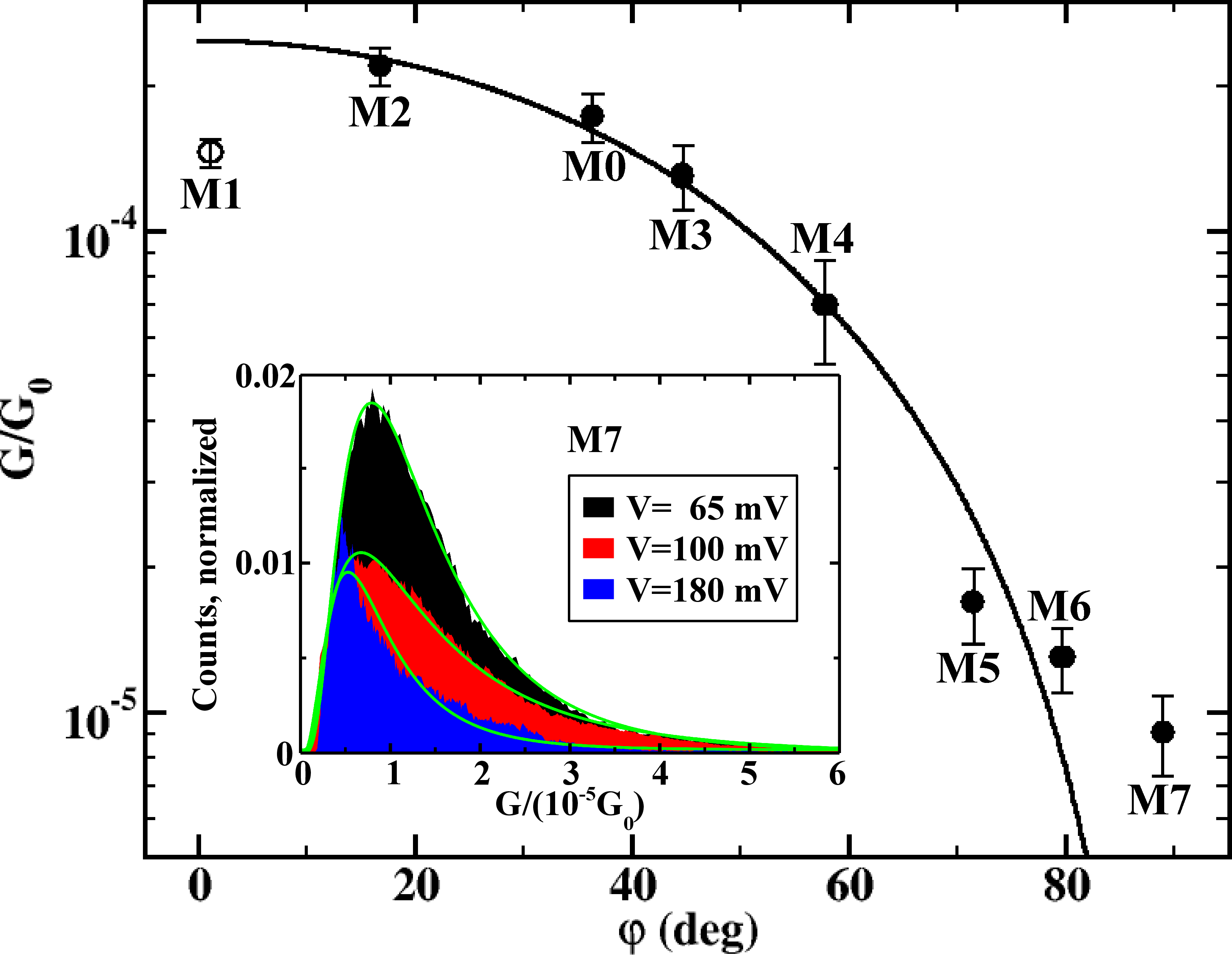} \caption{\label{fig:Gphi_exp}(Color online) Dependence of the experimental
conductance $G$ (points) on the torsion angle $\varphi$ for M1-M7
with $\varphi$ determined from X-ray structures. The solid line represents
a fit to $G=a\cos^{2}\varphi$ with $a=2.49\cdot10^{-4}G_{0}$. The
inset shows conductance histograms for M7, obtained at three different
bias voltages. The typical single-molecule conductance (points in
the main panel) is obtained as the average of the peak positions of
the fitted log-normal distributions at the three different biases,
and error bars are determined from the peak variations. For M7 this
yields $G=(9\pm2)\cdot10^{-6}G_{0}$.}
\end{figure}

We studied the conductance of single-molecule junctions by means of
a scanning-tunneling-microscopy-break-junction method (see Ref.\ \onlinecite{Mishchenko2010}).
The most probable or {}``typical'' molecular junction conductance
is determined as an average of the peak values in the conductance
histograms, measured at three different bias voltages (see Fig.~\ref{fig:Gphi_exp},
inset).%

When we plot this typical conductance as a function of the torsion
angle on a linear scale, we find a $G=a\cos^{2}\varphi$ dependence
with $a=2.49\cdot10^{-4}G_{0}$, as expected for off-resonant $\pi$-dominated
charge transport.\cite{Pauly2008b,Mishchenko2010} Here, $G_{0}=2e^{2}/h$
is the conductance quantum. Fig.~\ref{fig:Gphi_exp} shows a semi-logarithmic
plot of $G$ vs.\ $\varphi$. The graph reveals several distinct
features. The conductance of M1 is lower than expected from the general
trend. We observed a similar exceptional behavior of M1 in a recent
investigation of cyano-terminated molecules (i.e.\ S is replaced
by CN in Fig.~1),\cite{Mishchenko2011} showing, however, a higher
value than expected, and these irregularities are currently of an
unclear origin. On the other hand, there are systematic deviations
from the $\cos^{2}\varphi$ law for the molecules M6 and M7 with $\varphi\gtrsim80^{\circ}$.
A simple $\pi$-orbital model\cite{Viljas2008,Pauly2008a} looses
its validity for large torsion angles, and any other than $\pi$-$\pi$
couplings prevent the complete suppression of transport. We show below
that the residual couplings are of $\pi$-$\sigma$ type.

We observe that the conductance histograms, such as those shown in
the inset of Fig.~\ref{fig:Gphi_exp}, are asymmetric with a long
tail towards higher conductance values. The broad tail region could
be related to junctions with multiple molecules, modifications in
substrate-adsorbate coordination from junction to junction, atomic
rearrangements upon stretching, local surface roughness, or electrode-induced
changes of the average torsion angle due charge transfer and geometric
constraints. In contrast to results for biphenyldiamines,\cite{Venkataraman2006}
we do not observe clear correlations between the full-widths-at-half-maximum
of the conductance peaks and the expected differences in torsion-angle-related
energy barriers of the various BPDTs studied.\cite{Pauly2008b,Mishchenko2011}
If a substantial part of the experimental conductance scatter would
be due to the variation of the torsion angle of the biphenyl core,
then M0 with a low energy barrier for ring rotation of about 0.1 eV\cite{Pauly2008b}
should exhibit a particularly broad conductance distribution. However,
we see no evidence to support this hypothesis. In agreement with conclusions
from other works\cite{Malen2009a} we propose that the use of thiol
anchoring groups leads to the variation of the single-molecule conductance
being dominated by changes in the metal-molecule contact.

\section{Theoretical Procedures\label{sec:Methods}}

\subsection{Electronic structure and geometry optimization\label{sub:ElStruct_GeoOpt}}

Electronic structure calculations and geometry optimizations are performed
within DFT. We use the quantum chemistry package \noun{TURBOMOLE}
6.2.\cite{Ahlrichs1989} For all calculations, we employ the standard
basis set, def-SV(P), which is of split-valence quality with polarization
functions on all non-hydrogen atoms.\cite{Schafer:1994,Eichkorn1995,Eichkorn1997}
We treat all molecules and contact geometries as open-shell systems
with no unpaired electrons, and use BP86 as the exchange-correlation
functional.\cite{Becke1988,Perdew1986} Total energies are converged
to a precision of better than $10^{-6}$ a.u. and geometry optimizations
are carried out until the change of the maximum norm of the Cartesian
gradient is below $10^{-4}$ a.u.

\subsection{Charge transport calculations}

We determine conduction properties within the Landauer-B\"uttiker
formalism.\cite{Datta1997book} The energy-dependent transmission
$\tau(E)$ is expressed using standard Green's function techniques
(see also Appendix \ref{sec:Teigchans}). The conductance at low temperatures
is then given by

\begin{equation}
G=G_{0}\tau(E_{F})=G_{0}\sum_{n}\tau_{n}(E_{F}),\label{eq:G_Landauer}\end{equation}
 with $\tau_{n}$ being the transmission probability of the transmission
eigenchannel $n$.

\begin{figure}
\begin{centering}
\includegraphics[width=0.7\columnwidth]{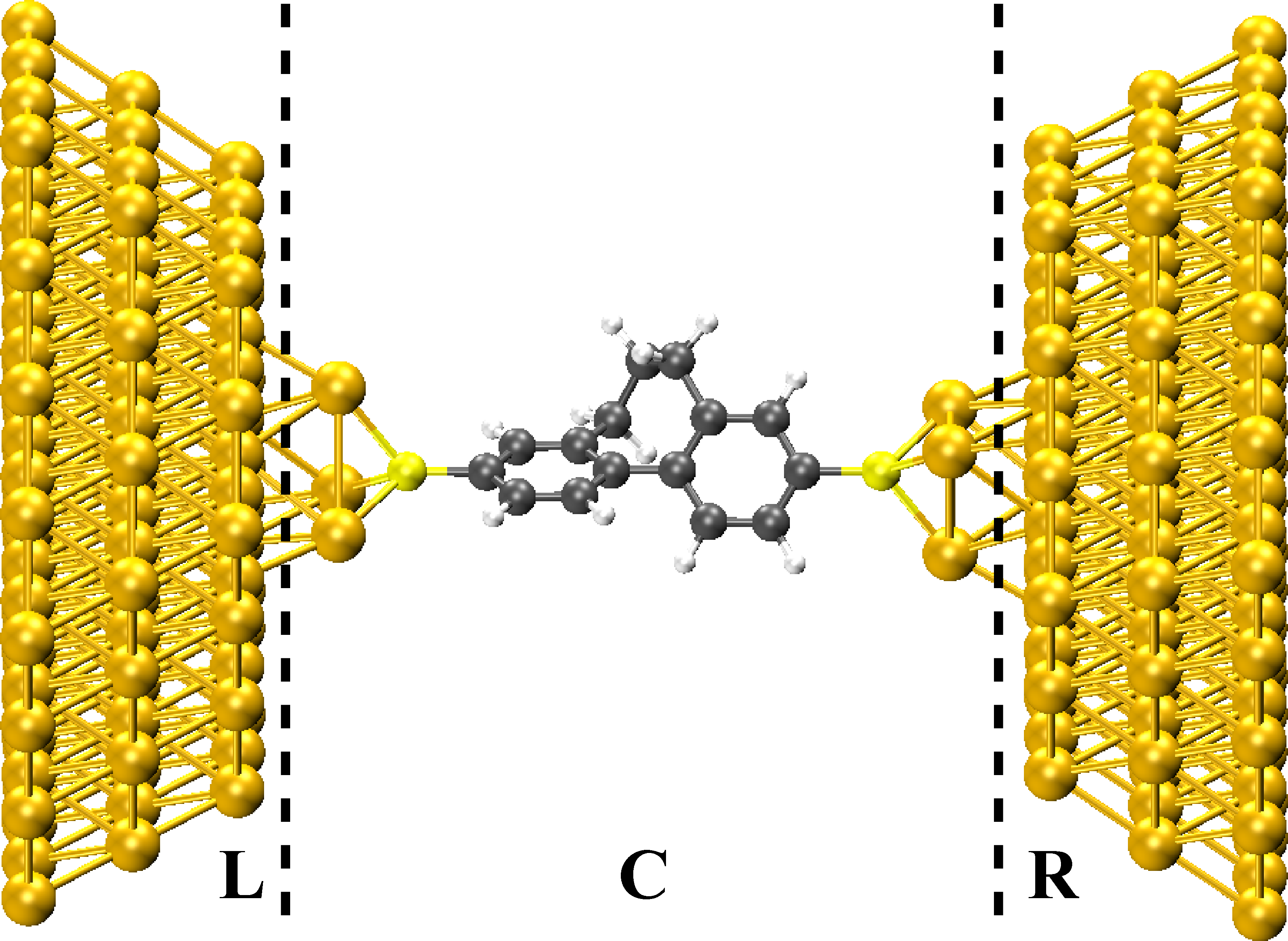} 
\par\end{centering}

\caption{\label{fig:bigsystem}(Color online) Division of the ECC into the
$L$, $C$, and $R$ regions. A large number of gold atoms (around
120 in $L$ and $R$, respectively) is used to represent the electrodes
in the DFT calculations.}
\end{figure}

In the calculations, we model the electrodes of a molecular junction
as perfect semi-infinite crystals to the left and to the right. The
molecule is connected to their surface by atomically sharp metal tips.
We describe this by computing a finite {}``extended central cluster''
(ECC), as displayed in Fig.\ \ref{fig:bigsystem}, into which large
parts of the metal electrodes are included to ensure the proper alignment
of molecular levels with respect to $E_{F}$.%

Due to the locality of the Gaussian basis sets employed, we are able
to partition the ECC into three subsystems, formed from basis states
in the left ($L$), central ($C$), and right ($R$) parts. The atoms
in the $L$ and $R$ regions of the ECC are assumed to represent that
part of the semi-infinite crystal surface which couples to $C$. We
extract the parameters for a description of region $C$ and its coupling
to the left and right electrode surfaces from the electronic structure
of the ECC. On the other hand, the surface Green's functions of the
$L$ or $R$ electrodes are constructed using parameters obtained
from a spherical Au cluster of several hundred atoms. This calculation
yields a Fermi energy of $E_{F}=-5.0$ eV. With these ingredients,
we compute the transmission probability $\tau(E)$. A more detailed
description of our cluster-based density-functional approach to quantum
transport can be found in Ref.\ \onlinecite{Pauly2008}.

In transport experiments with single molecules, often only the low-bias
conductance, proportional to the sum of the $\tau_{n}$ in Eq.~\eqref{eq:G_Landauer},
is measured. However, also the individual $\tau_{n}$ can be resolved.\cite{Scheer1998,Kiguchi2008}
On the theory side, in addition to the transmission probabilities
of the conduction eigenchannels, also the projection of their wavefunction
onto the central region can be obtained from quantities at hand in
the Green's-function formalism.\cite{Bagrets2007,Paulsson2007} In
order to construct energy-normalized transmission eigenchannel wavefunctions,
which can be compared to each other, we proceed along the lines of
Ref.\ \onlinecite{Paulsson2007}. Our efficient procedure, which
avoids the L\"owdin transformation of Ref.\ \onlinecite{Paulsson2007},
is presented in detail in Appendix~\ref{sec:Teigchans}.

\subsection{Contact geometries\label{sub:ContactGeo}}

\begin{figure}[t]
\begin{centering}
\includegraphics[width=0.85\columnwidth]{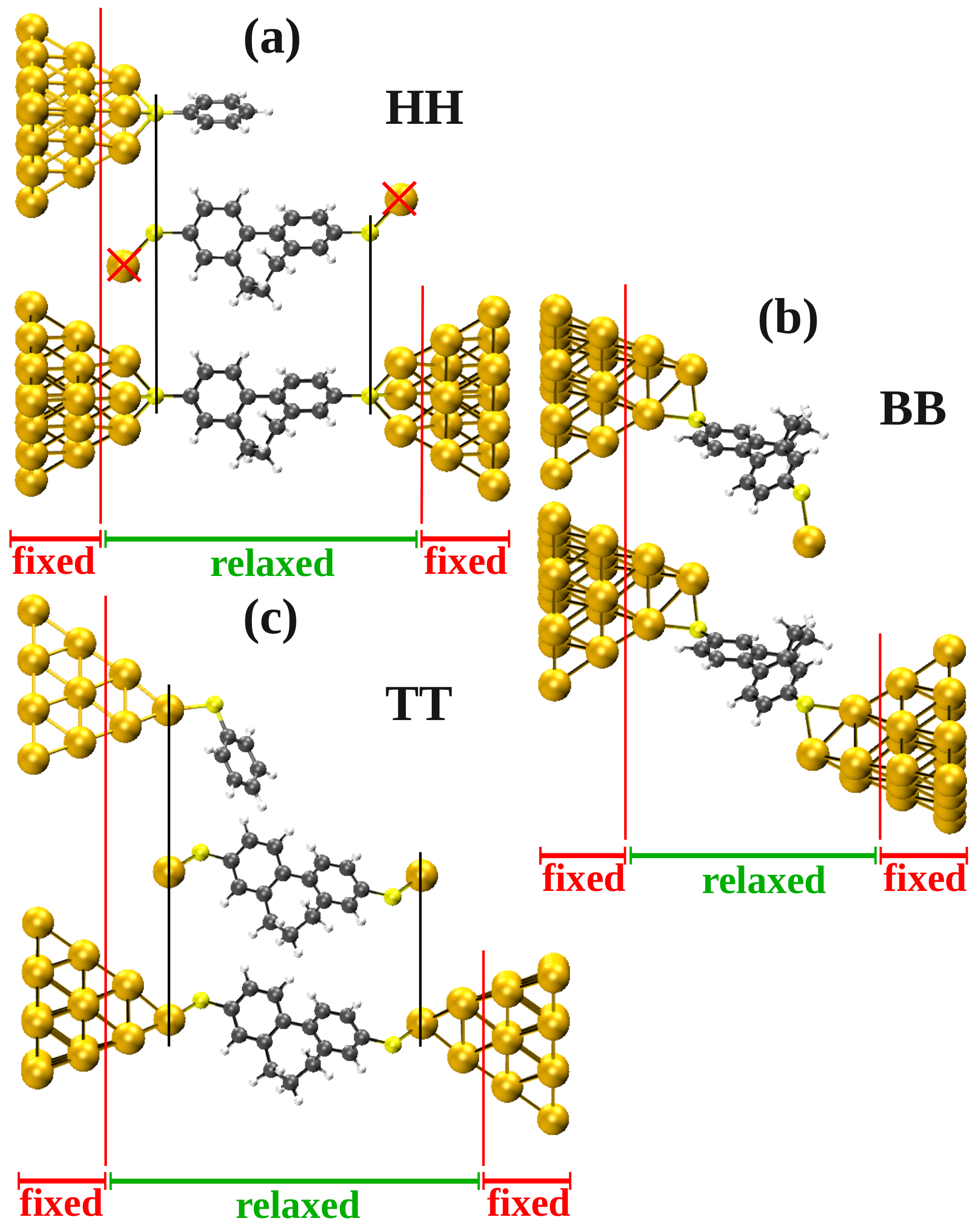} 
\par\end{centering}

\caption{\label{fig:geo_construction}(Color online) Procedure used to set
up the contact geometries for (a) HH, (b) BB, and (c) TT binding,
respectively. }
\end{figure}

The statistical nature of the single-molecule conductance experiments
(see Sec.\ \ref{sec:Experiments}) does not provide an a priori assignment
of representative junction geometries. Therefore, we have decided
to study three contact structures with different coordinations of
the terminal sulfur atoms. The procedure adopted to determine the
structure of the ECC is summarized in Fig.\ \ref{fig:geo_construction}.%

In order to model the molecular junctions, we connect the molecule
to two Au $\left\langle 111\right\rangle $ pyramids, both stemming
from the same ideal fcc Bravais lattice. We consider the following
three types: For hollow-hollow (HH) {[}Fig.\ \ref{fig:geo_construction}(a){]}
the S atoms of the molecule are bound at each side to three Au atoms,
for bridge-bridge (BB) {[}Fig.\ \ref{fig:geo_construction}(b){]}
to two Au atoms, and for top-top (TT) {[}Fig.\ \ref{fig:geo_construction}(c){]}
to only a single Au atom. Junctions of the form HB, HT etc.\ should
also occur in the experiments, but they are not considered here. 

For the determination of the HH and TT geometries, we start from the
gas-phase structure of each molecule (with SR=H in Fig.~\ref{fig:chemstruct}),
replace the terminal H atoms by an $\mbox{S-Au}_{1}$ group (R=Au$_{1}$
in Fig.~\ref{fig:chemstruct}), and compute ground-state geometries.
For HH contacts, the Au$_{1}$ atoms are removed. An Au$_{19}$ cluster,
resembling a Au $\left\langle 111\right\rangle $ pyramid with a thiolated
benzene attached, is computed separately. The cluster is positioned
at each side of the BPDT such that the S atoms on top of the pyramids
coincide with the S atoms of the molecule {[}Fig.~\ref{fig:geo_construction}(a){]}.
To obtain the TT geometries, the molecule is oriented such that each
Au$_{1}$ atom coincides with a tip atom of the Au$_{20}$ pyramids
{[}Fig.\ \ref{fig:geo_construction}(c){]}. To determine the equilibrium
structure for both HH and TT, the inner part is relaxed and only the
two outermost gold layers, consisting of $6$ and $10$ atoms, are
kept fixed in the ideal Au fcc structure.

For the BB geometries we follow slightly different steps. First the
terminal H atoms of the gas-phase molecules (with SR=H in Fig.~\ref{fig:chemstruct})
are replaced with S, one side is connected to a Au$_{20}$ pyramid
in bridge position, while the other one is terminated with Au$_{1}$.
The outermost gold layers of the pyramid are kept fixed, while the
rest is optimized. A second Au$_{20}$ pyramid is finally added to
the Au$_{1}$-terminated side, where the relative distances of the
binding S atom with respect to the new Au$_{20}$ cluster are chosen
to be the same as for the S atom in bridge position at the Au$_{20}$-terminated
side {[}Fig.~\ref{fig:geo_construction}(b){]}. Fixing again only
the two outermost Au layers, the structure is optimized to determine
the ground-state geometry.

We note that the contact geometries do not only differ with respect
to the coordination of the sulfur atoms to the gold electrodes, but
also in the stress exerted on the molecules. As visible in Fig.\ \ref{fig:geo_construction}(a),
the $\left\langle 111\right\rangle $ direction is located in the
ring plane of a mono-thiolated benzene molecule on top of a Au pyramid.
Since the S-S axis is along the same direction, the BPDT molecule
is expected to adopt a minimum-energy configuration inside the HH
junction with $\varphi$ close to its gas-phase angle. In contrast,
in the TT geometries the biphenyl derivative bridges the gold tip
atoms, which are opposite to each other. In this case the sulfur atoms
are deflected from their equilibrium positions, which would be located
along the $\left\langle 111\right\rangle $ direction on top of the
pyramids {[}Fig.\ \ref{fig:geo_construction}(c){]}. In the geometry
optimizations we find that the phenyl ring planes of the biphenyl
molecules tend to align parallel to the surfaces of the pyramids.
Since this may not be possible on both sides of the junction, some
torque is exerted. Beside effects related to charge transfer, which
may also be present for the HH contacts, geometric constraints thus
yield an additional contribution to the change of the torsion angle.
Similar effects as for TT are also present for the BB contacts, since
the orientations of the phenyl rings with respect to the gold pyramids
on both sides are generally different according to our construction.

We have determined binding energies by subtracting the total energy
of the contact geometries from those of the frozen separate parts,
namely the left and right Au clusters and the S-terminated biphenyl
(without hydrogen on the sulfur atoms). With this procedure, we find
the following averaged binding energies for the set of molecules:
$5.9\pm0.3$ eV (HH), $2.9\pm0.2$ eV (BB), and $2.2\pm0.2$ eV (TT).
Hence, we find a trend of decreasing binding energies with decreasing
coordination of the sulfur atoms to Au.

For reasons of computational feasibility, the structural optimizations
(and calculations of binding energies) are carried out with Au pyramids
consisting of $19$ atoms for HH, and $20$ atoms for TT and BB, respectively.
To ensure a proper description of the Fermi-level alignment in the
transport calculations, the gold pyramids are extended to $115$ (HH)
and $116$ atoms (BB, TT), as displayed in Fig.~\ref{fig:bigsystem}.
All added atoms are positioned on the ideal fcc lattice with a lattice
constant $a=0.408$ nm, matching those of the fixed layers for the
smaller pyramids. No further geometry optimization is carried out
for contacts with extended Au pyramids, and transport properties are
computed after a self-consistent, single-point DFT calculation.

\section{Theoretical Results\label{sec:Results}}

In this section we discuss in detail the effect of different contact
geometries on the molecular conformation of the BPDTs and their conduction
properties.

\subsection{Molecular Conformation\label{sub:Molecular-Conformation}}

\begin{figure}[t]
\begin{centering}
\includegraphics[width=1\columnwidth]{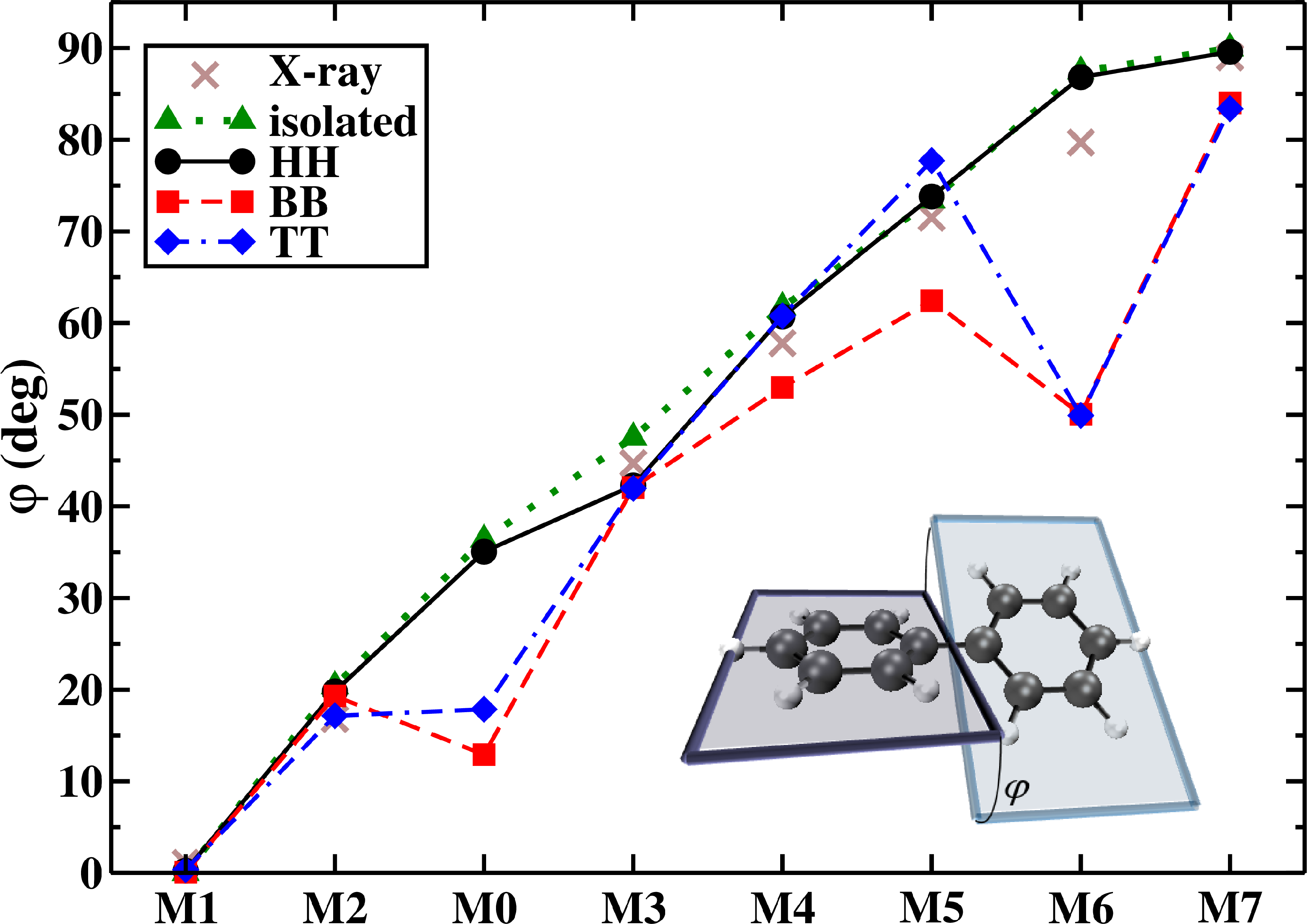} 
\par\end{centering}

\caption{\label{fig:phi_all}(Color online) Comparison of the torsion angle
$\varphi$ as determined from X-ray measurements and from DFT calculations
in the gas phase (isolated; SR=H in Fig.~\ref{fig:chemstruct}) as
well as in the molecular junctions (HH, BB, TT; see Fig.\ \ref{fig:geo_construction}).}
\end{figure}

Fig.\ \ref{fig:phi_all} shows the torsion angle between the two
phenyl rings for the molecules as determined by X-ray measurements
and by DFT calculations in the gas-phase as well as in the junction
geometries.%
 We notice that gas-phase angles (with SR=H in Fig.~\ref{fig:chemstruct})
generally coincide well with the angles from the X-ray measurements.\cite{Shaporenko2006,Vonlanthen2009,Vonlanthen2010}
The discrepancy for M6 by roughly $10^{\circ}$ has been observed
previously.\cite{Mishchenko2010} It is likely due to differences
between gas-phase and crystal structures caused by the limited stabilization
of the conformation, when there is just a single methyl group on each
phenyl ring. Note that no X-ray structure measurement exists for M0. 

For the contacted molecules deviations of $\varphi$ from the gas-phase
conformation are small for HH, but can be larger for the BB and TT
geometries. This is expected from the discussion in Sec.~\ref{sub:ContactGeo}.
The conformation of the alkyl-bridged BPDTs M1 to M5 is very stable.
A slight trend of increasing $\varphi$ variations for the molecules
with the longer, configurationally more flexible alkyl chains can
be recognized, however. The torsion angles of M0 and M6 result from
the balance between conjugation and modest steric repulsion effects
due to H atoms or single CH$_{3}$ groups in the ortho position with
respect to the ring-connecting carbons.\cite{Pauly2008b} Therefore,
their $\varphi$ should be rather sensitive to the geometric constraints
in the contacts or the charge transfer between the molecule and the
electrodes. As a result, deflections of $\varphi$ from the gas phase
values of up to $40^{\circ}$ occur in the calculations. In contrast,
the additional methyl side groups in M7 efficiently stabilize $\varphi$.\cite{Pauly2008b}

\subsection{Conductance}

\begin{figure}
\begin{centering}
\includegraphics[width=1\columnwidth]{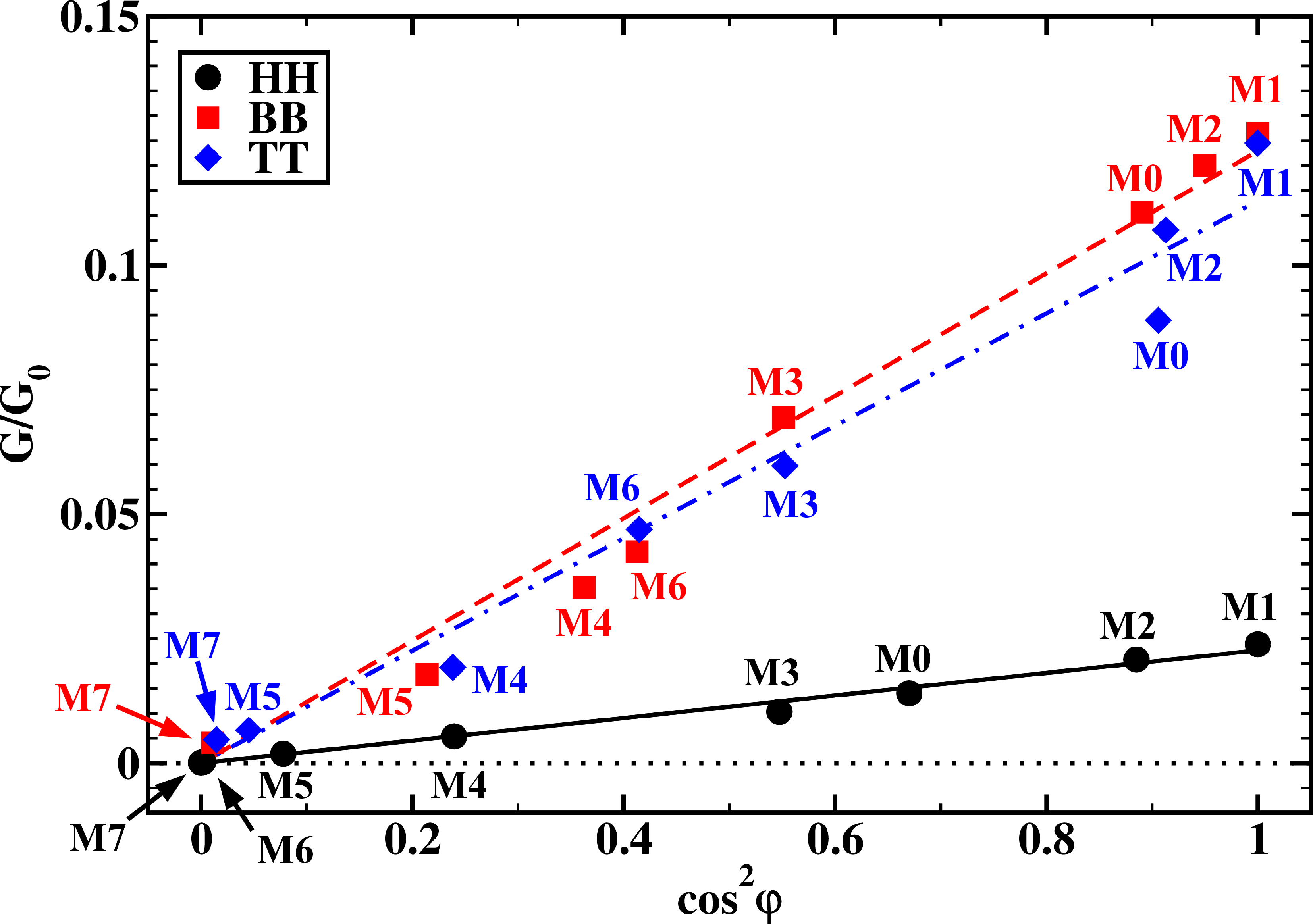} 
\par\end{centering}

\caption{\label{fig:Gcos2phi_all}(Color online) Calculated conductance as
a function of $\cos^{2}\varphi$ for the three types of contact geometries
HH, BB, and TT. The angles $\varphi$ are those of the computed junction
geometries. In all cases, lines show best fits for $G=a\cos^{2}\varphi$
with $a_{\mathrm{HH}}=2.3\cdot10^{-2}G_{0}$ (solid), $a_{\mathrm{BB}}=1.2\cdot10^{-1}G_{0}$
(dashed), $a_{\mathrm{TT}}=1.1\cdot10^{-1}G_{0}$ (dash-dotted).}
\end{figure}

\begin{figure}[!t]
\begin{centering}
\includegraphics[width=0.78\columnwidth]{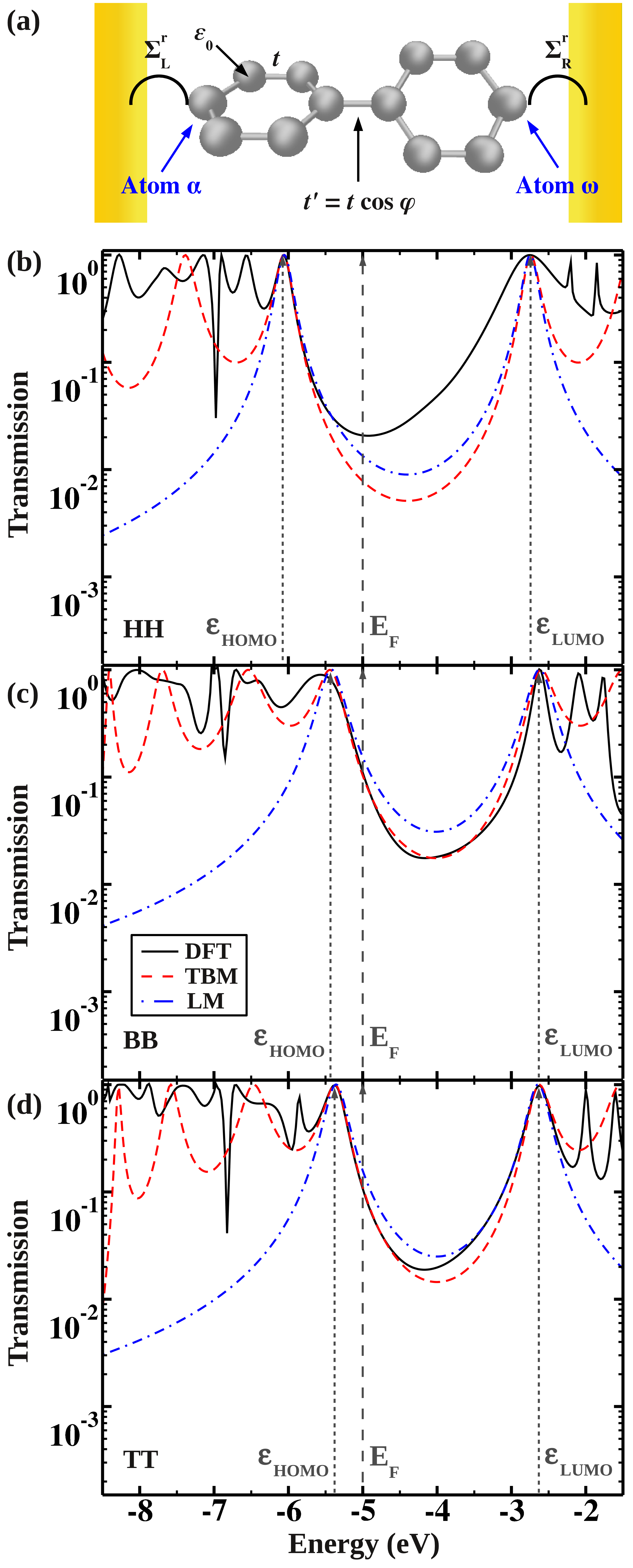}
\par\end{centering}

\caption{\label{fig:TE_TBM_LM}(Color online) (a) Sketch of the $\pi$-orbital
TBM used to describe the transport through BPDT molecules. $\epsilon_{0}$
is the onsite energy, identical for all carbon atoms, $t$ the coupling
between nearest-neighbor atoms on each phenyl ring, $t'=t\cos\varphi$
the inter-ring coupling, and $\varphi$ the torsion angle realized
in the particular junction geometry. The terminal atoms of the biphenyl
backbone to the left and right are indexed $\alpha$ and $\omega$,
respectively, and their couplings to the $L,R$ electrodes are described
by the self energies $\Sigma_{L}^{r},\Sigma_{R}^{r}$. (b-d) Transmission
as a function of energy for M2 in the different junction geometries
HH, BB, and TT. The solid line is the DFT result, the dashed line
the fi{}t with the TBM, and the dash-dotted line corresponds to the
LM. Vertical dashed lines indicate, in the order of increasing energy,
the position of the HOMO, the Fermi energy, and the LUMO.}
\end{figure}

In Fig.\ \ref{fig:Gcos2phi_all} we present the computed conductance
values as a function of the torsion angle $\varphi$, which the biphenyl
molecules adopt in the optimized junction geometries.%
 On the linear conductance scale we find a reasonable $G=a\cos^{2}\varphi$
dependence for all binding situations with best fit coefficients%
\footnote{We note that small differences in the slope values $a$ as compared
to Ref.~\onlinecite{Mishchenko2010} arise from the study of a different
junction geometry for TT and a slightly modified ECC for BB.%
} $a_{\mathrm{HH}}=2.3\cdot10^{-2}G_{0}$, $a_{\mathrm{BB}}=1.2\cdot10^{-1}G_{0}$,
$a_{\mathrm{TT}}=1.1\cdot10^{-1}G_{0}$. This behavior is characteristic
for off-resonant charge transport dominated by $\pi$-$\pi$ coupling
and is consistent with the experimental observations. Fig.~\ref{fig:TE_TBM_LM}
shows, for the sample molecule M2, that irrespective of the coordination
site, the transport is indeed off-resonant and dominated by the highest
occupied molecular orbital (HOMO) level.%

The results suggest large variations in the conduction properties
for the different coordination sites of sulfur to gold. The conductance
of junctions with HH geometry is roughly one order of magnitude lower
as compared to BB and TT, with the sequence of slopes $a_{\mathrm{HH}}\ll a_{\mathrm{TT}}\approx a_{\mathrm{BB}}$.
Similar behavior of the conductance of dithiolated aromatic molecules
on binding site has been reported before by other authors.\cite{Xue2003,Tomfohr2004,Li2007} 

For aliphatic alkane molecules, the conductance in the bridge-bonded
configuration was reported to be higher than in the top-bonded one.\cite{Li2008}
While these findings are compatible with our results in Fig.~\ref{fig:Gcos2phi_all},
transport through alkanes is $\sigma$-like\cite{Li2008,Arroyo2010,Kim2011}
and hence differs substantially from the typical $\pi$-dominated
transport through aromatic molecules. Beside the coordination of the
anchoring group the molecular tilt, which determines the overlap of
the delocalized $\pi$ electrons with the electrode, hence plays a
crucial role for the conductance of aromatic molecules.\cite{Haiss2008,Quek2009,Diez-Perez2011}
We discuss these aspects further below. However, we note that our
contact geometries do not allow us to clearly separate the effects
of coordination site and tilt, since both are changed simultaneously.

With regard to absolute values, we observe that the calculated conductances
are three (BB and TT geometries) and two (HH geometry) orders of magnitude
higher than the experimental ones.\cite{Vonlanthen2009,Mishchenko2010}
We attribute this overestimation mostly to the interpretation of Kohn-Sham
orbitals as approximate quasi-particle energies.\cite{Quek2007,Strange2011}
However, also the experimentally measured conductances are subject
to uncertainties. Indeed, we compare our results to the {}``typical''
experimental values, as given by the peak position in a conductance
histogram, and these peaks are rather broad. Further variations of
molecular conductance, for example due to interactions of the molecules
with the solvent and other close-by biphenyl molecules or the influence
of vibrations, have not been accounted for in our calculations of
static junctions in vacuum. The differences on a quantitative level
remain as a major challenge for future work.

Finally, we note that our calculations do not reproduce the experimental
deviation observed for M1. \cite{Mishchenko2010} Since $\varphi$
is unchanged upon contacting (see Fig.\ \ref{fig:phi_all}), we can
exclude an explanation based on conformational changes, which would
decrease the degree of conjugation and lead to reduced conductances.
In spite of the slightly bent structure due to the short CH$_{2}$
bridge (see Fig.~\ref{fig:chemstruct}), the intact M1 shows the
highest calculated conductances for all coordination geometries (see
Fig.~\ref{fig:Gcos2phi_all}).

\subsection{Analysis of transmission resonances}

In order to understand better the charge transport through the BPDT
single-molecule junctions, we analyze the transmission in terms of
a TBM and a LM. We use the TBM of Ref.~\onlinecite{Viljas2008},
which is sketched in Fig.~\ref{fig:TE_TBM_LM}(a). It describes the
delocalized $\pi$-electron system, relevant for transport away from
the perpendicular orientation of the phenyl rings. The H\"uckel-like,
molecular Hamiltonian contains three parameters, namely the onsite
energy $\epsilon_{0}$ of each carbon atom, the hopping $t$ between
nearest-neighbor atoms on each ring, and the torsion angle $\varphi$,
specific to the considered junction geometry. Together, $t$ and $\varphi$
determine the matrix element between the ring-connecting carbon atoms
$t'=t\cos\varphi$. For the description of transport we make use of
the wide-band approximation, according to which the retarded self
energy $\Sigma_{X}^{r}$ due to the coupling to the electrode $X=L,R$
is determined by the line-broadening matrix $\Gamma_{X}$ as $\Sigma_{X}^{r}=-i\Gamma_{X}/2$
(cf.\ Eq.~\ref{eq:Gamma}). We assume a symmetric junction $\Gamma=(\Gamma_{L})_{\alpha\alpha}=(\Gamma_{R})_{\omega\omega}$
and, in line with the nearest-neighbor coupling in the molecule, consider
the self energy to be nonvanishing only on the terminal carbon atoms
$\alpha$ and $\omega$ of the biphenyl backbone {[}see Fig.~\ref{fig:TE_TBM_LM}(a){]}.
The TBM is hence characterized by the four parameters $\varphi,\epsilon_{0},t,\Gamma$,
where $\varphi$ is fixed by the considered junction geometry.

The parameters of the LM are derived from those of the TBM. For that
purpose we solve the non-Hermitian eigenvalue problem $\sum_{k}(H+\Sigma^{r})_{jk}v_{k}^{\mu}=\lambda_{\mu}v_{k}^{\mu}$
and select the complex eigenvalues corresponding to the HOMO and the
lowest unoccupied molecular orbital (LUMO). In the eigenvalue equation
$H_{jk}$ and $(\Sigma^{r})_{jk}=(\Sigma_{L}^{r}+\Sigma_{R}^{r})_{jk}$
are the Hamiltonian matrix and the self-energy matrix of the TBM,
respectively, and $\lambda_{\mu}=\epsilon_{\mu}+i\gamma_{\mu}$. We
measure the real part of the complex eigenvalues with respect to the
Fermi energy, introducing $\tilde{\epsilon}_{H}=\epsilon_{HOMO}-E_{F}$
and $\tilde{\epsilon}_{L}=\epsilon_{LUMO}-E_{F}$. Due to the symmetries
of the TBM we find for the imaginary parts $\gamma_{HOMO}=\gamma_{LUMO}$,
and set $\tilde{\Gamma}=|\gamma_{HOMO}|$. From the relation between
the TBM and the LM discussed in Appendix \ref{sec:Rel-TBM-LM} {[}see
Eq.~(\ref{eq:TE_LM}){]}, we can identify $\tilde{\Gamma}$ with
the width of the Lorentzian transmission resonances related to the
HOMO and the LUMO. Finally, we determine the transmission for the
LM via Eq.~(\ref{eq:TE_LM}) as a sum over these two frontier orbitals
only. The LM is thus characterized by $\tilde{\epsilon}_{H},\tilde{\epsilon}_{L},\tilde{\Gamma}$
and is specific to a certain molecule and junction geometry, as described
by $\varphi,\epsilon_{0},t,\Gamma$ in the TBM.

Using the physically motivated TBM, we have fitted the transmission
$\tau(E)$ of the well-conjugated molecules M1-M4, as determined by
the DFT calculations. Setting $\varphi$ to the value of the torsion
angle realized in the particular junction geometry, we place special
emphasis on a good fit in the region of the HOMO-LUMO gap. When such
a fit is too ambitious due to the simplicity of the TBM, we describe
well at least the region between the HOMO and $E_{F}$, to determine
effective parameters for the dominant transmission resonance, as well
as the position of the LUMO peak. In this way, we obtain the values
for $\epsilon_{0},t,\Gamma$ given in Table \ref{tab:TBM_LM_params}.%
\begin{table}[t]
\begin{centering}
\begin{tabular}{c|c|c|c|c|c|c}
 & $\epsilon_{0}$  & $t$  & $\Gamma$  & $\text{\ensuremath{\tilde{\epsilon}}}_{H}^{M2}$  & $\text{\ensuremath{\tilde{\epsilon}}}_{L}^{M2}$  & $\tilde{\Gamma}^{M2}$\tabularnewline
\hline
\hline 
HH  & -4.40  & -2.30  & 0.70  & -1.05  & 2.25  & 0.11\tabularnewline
BB  & -4.02  & -1.95  & 1.10  & -0.42 & 2.38  & 0.18\tabularnewline
TT  & -4.00  & -1.90  & 0.96  & -0.36  & 2.36 & 0.15\tabularnewline
\end{tabular}
\par\end{centering}

\caption{\label{tab:TBM_LM_params}Parameters of the TBM $\epsilon_{0},t,\Gamma$
obtained by fitting the DFT-based $\tau(E)$ curves for M1-M4. The
parameters $\tilde{\epsilon}_{H}^{M2},\tilde{\epsilon}_{L}^{M2},\tilde{\Gamma}^{M2}$
of the LM are those derived from the TBM for M2. All values are given
in units of eV.}
\end{table}
 Specific LM parameters for M2 are provided in the same table, and
DFT, TBM, and LM transmission curves for M2 are shown in Fig.~\ref{fig:TE_TBM_LM}(b-d).

The differences between the curves of the TBM and the LM in Fig.~\ref{fig:TE_TBM_LM}(b-d)
in the region of the HOMO-LUMO gap illustrate approximations related
to the neglect of interference effects in the LM. Indeed, we find
that the transmission is slightly overestimated when it is regarded
as the superposition of incoherent transmission resonances. In the
following we restrict our discussion to the parameters of the LM for
M2, since they are easy to interpret and those of the generic TBM
contain similar information.

The data in Table \ref{tab:TBM_LM_params} shows very similar values
of $\tilde{\Gamma}^{M2}$ for the different junction geometries. While
the increasing linewidth $\tilde{\Gamma}^{M2}$ when going from TT
to BB is consistent with the expectation of a better electronic coupling
for a higher coordination of the sulfur atom, also the molecular tilt
plays a role. The perpendicular orientation of the BPDTs for geometry
HH thus leads to a reduced $\tilde{\Gamma}^{M2}$. As an important
conclusion, the values of $\tilde{\epsilon}_{H}^{M2}$ and $\tilde{\epsilon}_{L}^{M2}$
show that the HOMO is closer to $E_{F}$ than the LUMO by more than
$1$ eV. In addition, the reduced conductance for HH in Fig.~\ref{fig:Gcos2phi_all}
is explained by the HOMO level being around 0.5 eV further away from
$E_{F}$ than for BB and TT.

We attribute the shift of the HOMO level towards lower energies for
increasing coordination number of the sulfur atoms to the different
amounts of transferred charge at the molecule-Au interface. Indeed,
both L\"owdin and electrostatic-potential-derived charges yield a
leakage of electrons from the molecule, including the S atoms, to
the Au electrodes, when going from TT over BB to the HH geometry.
Variations of the conductance therefore mostly arise from changes
in the alignment of the HOMO level with respect to the Fermi energy
of the Au electrodes, and originate from charge redistributions, which
are sensitive to the coordination site of the sulfur atom at the molecule-electrode
interface.

\subsection{Transmission eigenchannels}

\begin{figure}[!t]
\begin{centering}
\includegraphics[width=1\columnwidth]{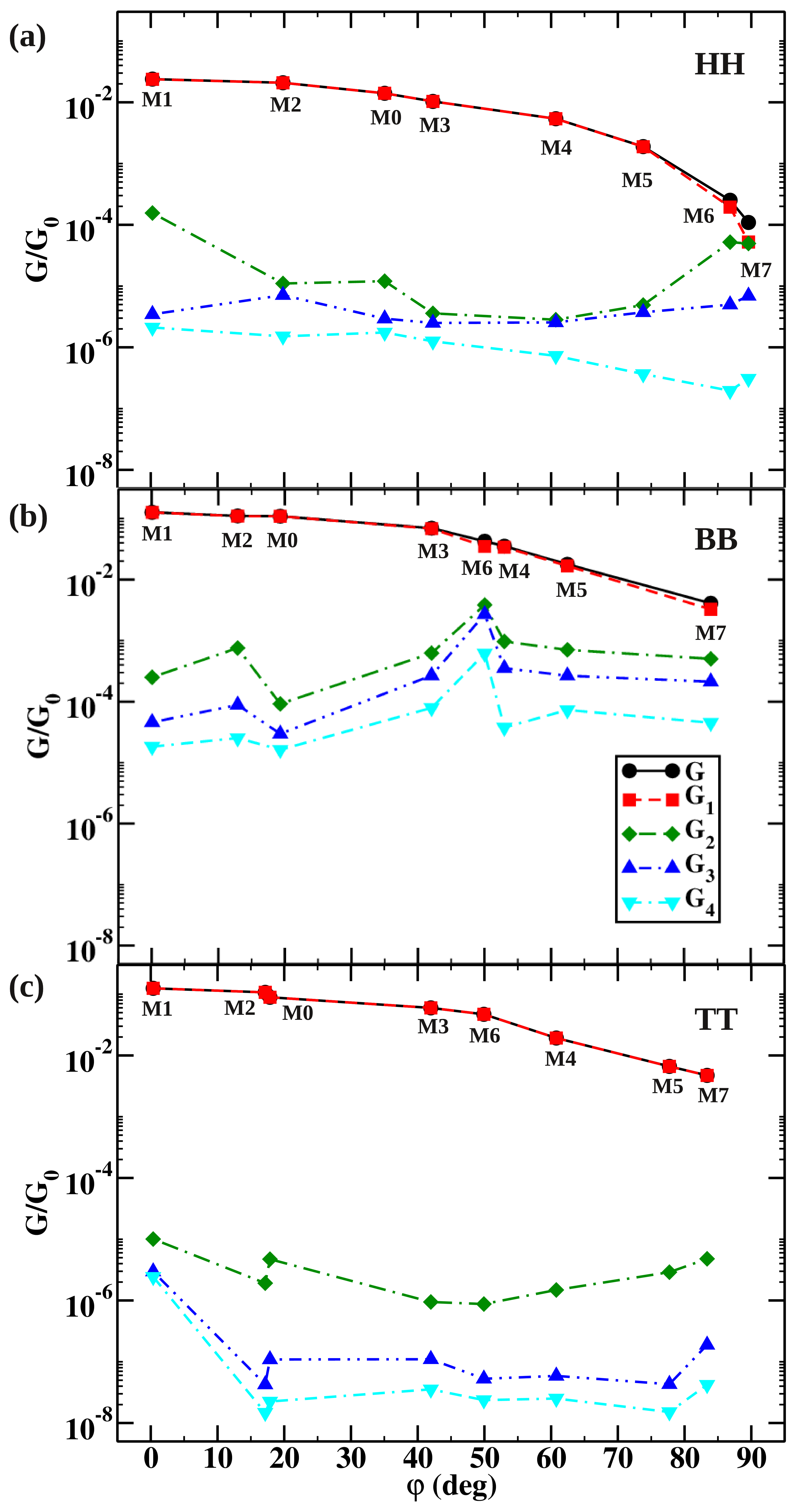} 
\par\end{centering}

\caption{\label{fig:channel_plots}(Color online) Calculated conductance $G$
and the conductance $G_{n}=G_{0}\tau_{n}(E_{F})$ with $n=1,\ldots,4$
of the four transmission eigenchannels with the highest contribution
to $G$ for the set of BPDTs in (a) HH, (b) BB, and (c) TT configurations.}
\end{figure}

\begin{figure*}
\begin{centering}
\includegraphics[width=1.6\columnwidth]{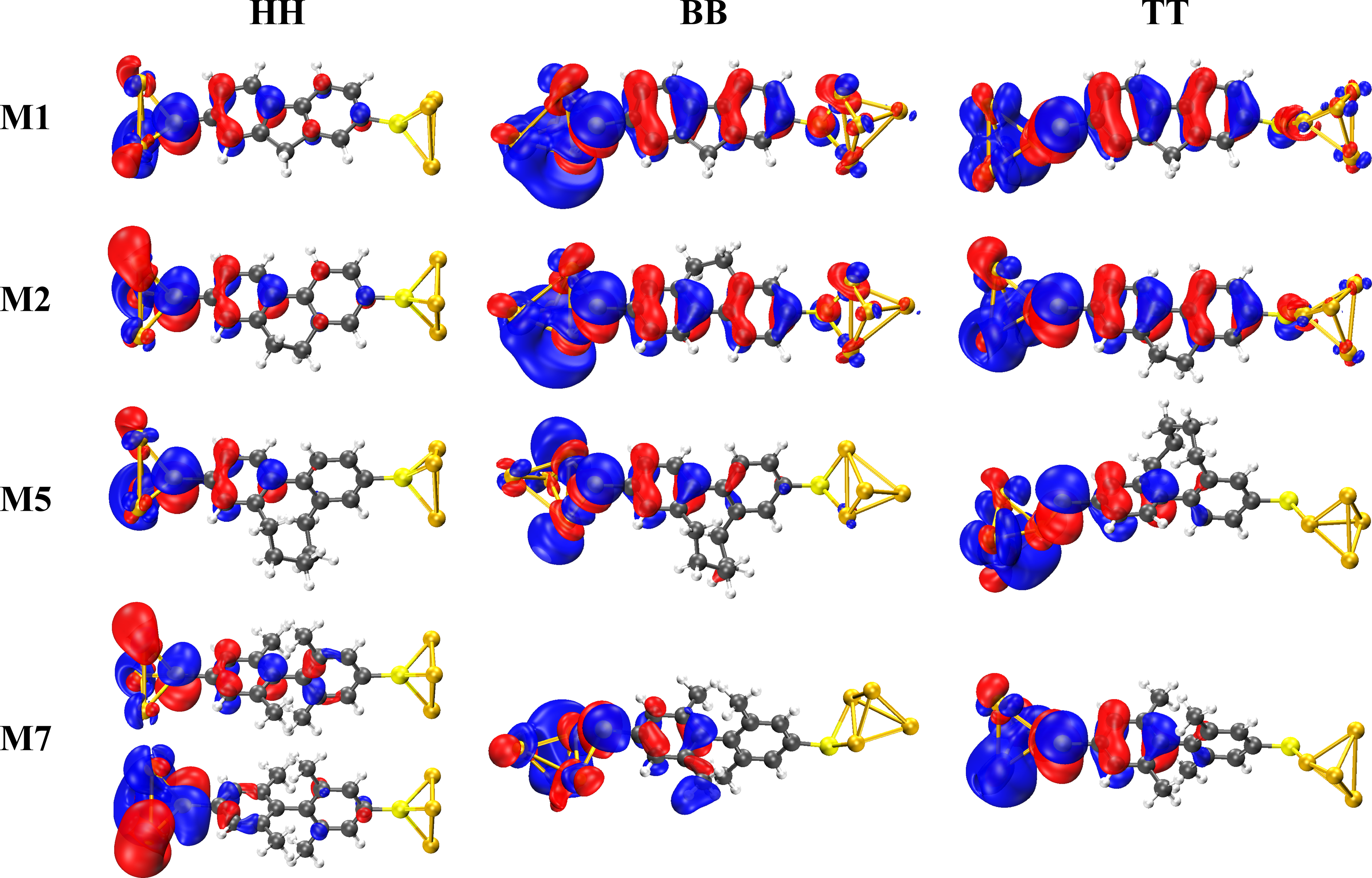} 
\par\end{centering}

\caption{\label{fig:wave-functions}(Color online) Wavefunction of the dominant,
left-incoming transmission eigenchannel for selected BPDT molecules
in the HH, BB, and TT geometries. The same isosurface value of the
wavefunctions is used in all the plots to allow for their comparison.
However, the isosurface value has been reduced by a factor of 4 on
the right phenyl ring of M7 for HH to visualize the $\pi$-$\sigma$
and $\sigma$-$\pi$ character of the two eigenchannel wavefunctions,
which yield the same contribution to the conductance.}
\end{figure*}

To explore further the electron transport through BPDT molecules,
especially for the situation $\varphi\simeq90^{\circ}$ where the
TBM looses its validity, we consider the eigenchannel decomposition
of the conductance and the corresponding wavefunctions. The results
are displayed in Figs.\ \ref{fig:channel_plots} and \ref{fig:wave-functions}.%

We observe one dominant eigenchannel, whose transmission probability
is decreasing gradually with increasing torsion angle for geometries
with $\varphi\lesssim80^{\circ}$ (Fig.\ \ref{fig:channel_plots}).
The wavefunction of this channel is formed from those $p$ orbitals
of the C atoms, which are perpendicular to the phenyl-ring planes
(see the results for M1 and M2 in Fig.~\ref{fig:wave-functions}).
The resulting $\pi$ orbitals, which comprise the terminal thiol groups,
hence exhibit nodes in the ring planes. The findings agree with the
expectation that for the small torsion angles, resulting in a high
degree of conjugation, electric transport should occur via the delocalized
$\pi$-electron system of the BPDTs. 

The $\pi$-$\pi$ coupling between the rings is suppressed for $\varphi\approx90^{\circ}$,
since it varies as $\cos\varphi$.\cite{Tomfohr2004,Pauly2008b,Pauly2008a}
In this case the molecular states become more localized on the individual
rings. The incoming Bloch waves from the leads can still couple through
the sulfur linker atom into the $\pi$-electron system of one of the
rings, but they are back-reflected at the ring-connecting carbon atom.
This results in a large suppression of the transmission (Fig.\ \ref{fig:channel_plots}),
and becomes manifest in a low amplitude of the wavefunction on the
second ring (see the results for M5 and M7 in Fig.~\ref{fig:wave-functions}).
In this regime, the $\pi$-$\sigma$ coupling, proportional to $\sin\varphi$,
dominates.\cite{Tomfohr2004,Pauly2008a} The $\sigma$ character of
the wavefunctions is apparent from the absence of nodal planes in
the phenyl ring planes and the high amplitude of the eigenchannel
wavefunction on the axis which connects the neighboring carbon atoms.

The isolated biphenyl molecules M0 and M7 (SR=H in Fig.~\ref{fig:chemstruct})
with $\varphi$ set to $90^{\circ}$ possess $D_{2d}$ symmetry. Then,
$\sigma$-$\pi$ and $\pi$-$\sigma$ orbitals are degenerate, which
should lead to two dominant transmission eigenchannels with the same
contribution to the conductance.\cite{Tomfohr2004,Pauly2008a} However,
the presence of the electrodes generally leads to a low symmetry of
the junction as a whole and may also modify the molecular geometry.
Hence, it is interesting to analyze the degeneracy of eigenchannels
in the different coordination geometries for M7 with the nearly perpendicular
gas phase torsion angle. Using the ratio of the channel conductances
$G_{2}/G_{1}$ with $G_{n}=G_{0}\tau_{n}(E_{F})$ as a measure for
the degeneracy, we find the values given in Table \ref{tab:chan_deg}.%
\begin{table}[t]
\begin{centering}
\begin{tabular}{c|c|c|c}
 & HH & BB & TT\tabularnewline
\hline
\hline 
M7  & $0.95$ & $0.15$ & $1.0\cdot10^{-3}$\tabularnewline
\end{tabular}
\par\end{centering}

\caption{\label{tab:chan_deg}Ratio $G_{2}/G_{1}$ of the highest eigenchannel
contributions to the conductance for M7 in the three junction geometries
studied.}
\end{table}

The data in Table \ref{tab:chan_deg} demonstrates the general absence
of the channel degeneracy and a high sensitivity of $G_{2}/G_{1}$
to the junction geometry. Only for the HH contact geometry we find
a nearly perfect degeneracy of the two dominant transmission eigenchannels.
Consistent with this, Fig.~\ref{fig:wave-functions} demonstrates
that their wavefunctions are indeed of $\pi$-$\sigma$ and $\sigma$-$\pi$
type. The degeneracy can be explained by the fact that M7 in the HH
geometry stands perpendicular to the electrodes. The torsion angle
of the contacted molecule is hence close to those in the gas phase
(see Fig.\ \ref{fig:phi_all}), and the overlap of the molecular
$\pi$ orbitals with the electrode states is such that the degeneracy
of molecular orbitals is not strongly lifted. Therefore, the transmission
reflects symmetry properties of the molecule. For the BB and TT junctions,
deviations from the channel degeneracy result from the geometric constraints
set by the electrodes, which cause $\varphi$ in the junctions to
deviate from $90^{\circ}$, and from the asymmetric overlap of the
molecular $\pi$ states with the electrode states to the left and
right. Thus, our results clearly show that the reduced symmetry of
the complete junction has to be considered for transport and not just
the symmetry of the isolated molecule alone.\cite{Solomon2006}

These findings suggest that measurements of the transmission eigenchannel
degeneracy may serve as a sensitive probe to determine the coordination
geometry in biphenyl-type single-molecule junctions. However, there
are several factors not included in our idealized treatment. Thus,
it would be interesting to study, how strongly a finite bias voltage
will lift an existing $\pi$-$\sigma$ and $\sigma$-$\pi$ channel
degeneracy by breaking of the left-right symmetry. Furthermore, also
dynamic effects due to vibrational modes and Jahn-Teller distortions
should lead to an effective splitting of the two dominant eigenchannels.
Beside these issues, it remains an experimental challenge to determine
the conduction eigenchannel transparencies for contacts with a low
transmission, since the existing techniques, employing superconducting
electrodes\cite{Scheer1998} or shot noise,\cite{Kiguchi2008} yield
very low signals in such situations.

Our calculations illustrate that the alkyl chains do not participate
significantly in transport, as expected from the large gaps between
HOMO and LUMO levels of alkanes.\cite{Li2008} Considering the dominant
transmission eigenchannels in Fig.\ \ref{fig:wave-functions}, we
see that there is indeed practically no weight of the wavefunction
on the alkyl chain, even for the short chains present in M1 and M2.

\section{conclusions\label{sec:Conclusions}}

Motivated by recent experiments,\cite{Vonlanthen2009,Mishchenko2010}
we have presented a detailed theoretical analysis of the charge transport
properties of Au-BPDT-Au single-molecule junctions. The three different
types of contact geometries in our DFT-based study differed in essential
aspects at the molecule-metal interface. They were mainly the coordination
site of the anchoring sulfur atoms and the tilt of the molecule with
respect to the electrodes. Given the extensive statistical analysis
in the experiments, this set of geometries is clearly very limited.
Furthermore, without an analysis of the junction formation process,
it is difficult to make a statement on the probability of their occurrence.
However, we hope that they can be used to describe general trends,
such as the influence of molecular conformation on conductance and
the variability of transport properties with contact geometry.

We have investigated electrode-induced changes of the molecular conformation
due to charge transfer and geometric constraints and find that they
are rather small for most molecules and types of junctions considered
here. Compared to the somewhat larger variations for M0 and M6, whose
$\varphi$ is not fixed by an alkyl strap or strong steric effects,
our calculations show that the appropriate design of the side groups
can help to stabilize the torsion angle.

The transport calculations confirm a $\cos^{2}\varphi$ dependence
of the conductance for the well-conjugated molecules in each type
of junction geometry. This is in accordance with the experimental
observations and is characteristic for off-resonant transport through
the $\pi$-electron system.\cite{Pauly2008b} For biphenyl molecules
with torsion angles close to the perpendicular orientation, however,
we observe systematic deviations in our experimental data from the
$\cos^{2}\varphi$ law predicted by a simplified $\pi$-orbital TBM.
In that regime of a broken conjugation, our analysis of DFT-based
transmission eigenchannel wavefunctions reveals residual conductance
contributions from a pair of $\pi$-$\sigma$-type conduction channels.

Finally, our calculations suggest that molecular junctions with sulfur
atoms bound to the {}``hollow'' site of gold electrodes could exhibit
an order of magnitude smaller conductance as compared to junctions
with sulfur atoms bound via {}``top'' or {}``bridge'' sites. Our
analysis shows that the transport is dominated by the molecular HOMO
level in all cases, and variations of the conductance arise from changes
in the alignment of that level with respect to the Fermi energy of
the Au electrodes. These changes in turn originate from the charge
transfer between the molecule and the electrode, which is sensitive
to the coordination site of the sulfur atom.
\begin{acknowledgments}
We acknowledge discussions with A.\ Bagrets, F.\ Evers, and V.\
Meded. R.\ Ahlrichs and M.\ Sierka are thanked for providing us
with TURBOMOLE. M.B.\ was supported through the DFG CFN (Project
C3.6) and the DFG SPP 1243, F.P.\ through the Young Investigator
Group, and J.K.V.\ through the Academy of Finland. D.V.\ and M.M.\
acknowledge funding by the Swiss National Science Foundation and the
Swiss National Center of Competence in Research {}``Nanoscale Science''.
The work of A.M.\ and T.W.\ was financed by the Swiss National Science
Foundation (200021.124643, NFP62), the ITN FP7 network FUNMOLS, the
DFG SPP 1243, and the University of Bern.
\end{acknowledgments}
\appendix

\section{Determination of transmission eigenchannels\label{sec:Teigchans}}

In this appendix, we provide further details on how we determine the
transmission eigenchannels, in particular their wavefunctions. The
result is equivalent to that of Ref.\ \onlinecite{Paulsson2007}.
However, our procedure avoids the L\"owdin orthogonalizations and
uses, instead, a consistent formulation in terms of nonorthogonal
basis states. This reduces the numerical effort and eliminates possible
numerical instabilities resulting from the forward and backward L\"owdin
transformations. Isosurfaces of eigenchannel wavefunctions employing
this scheme are plotted in Fig.\ \ref{fig:wave-functions}.

To compute the charge transport, we divide the nanocontact into a
left ($L$), central ($C$), and right ($R$) region (see Fig.\ \ref{fig:bigsystem})
and classify the states of our local, nonorthogonal basis $|e_{i}\rangle$
accordingly. The $C$ region is assumed to be long enough to neglect
the elements $H_{LR}=H_{RL}^{T}$ and $S_{LR}=S_{RL}^{T}$ of the
real and symmetric Hamiltonian $H_{jk}=\langle e_{j}|\breve{H}|e_{k}\rangle$
and overlap $S_{jk}=\langle e_{j}|e_{k}\rangle$. Here $\breve{H}$
is the Hamiltonian operator in the combined $L,C,R$ space.

Adopting a notation along the lines of Refs.\ \onlinecite{Paulsson2007,Pauly2008},
we express the energy-dependent transmission as \begin{equation}
\tau(E)=\mathrm{Tr}\left[A_{L}(E)\Gamma_{R}(E)\right],\label{eq:Transmission}\end{equation}
 where we define the spectral function \begin{equation}
A_{X}(E)=G_{CC}^{r}(E)\Gamma_{X}(E)G_{CC}^{a}(E).\label{eq:AX_def}\end{equation}
 Here and below, $X$ stands for either $L$ or $R$ ($X=L,R$). $A_{X}(E)$
is the contribution to the full spectral density of $C$ from scattering
states originating in lead $X$.\cite{Paulsson2007} In the expression,
\begin{equation}
G_{CC}^{r}(E)=\left[\left(E+i\eta\right)S_{CC}-H_{CC}-\Sigma_{L}^{r}(E)-\Sigma_{R}^{r}(E)\right]^{-1}\label{eq:GrCC}\end{equation}
 is the retarded Green's function of the device (or $C$ region),
with $\eta>0$ an infinitesimal constant, and $G_{CC}^{a}=(G_{CC}^{r})^{\dagger}$
is the advanced function. For $G_{CC}^{r}$ we need the self energies
\begin{equation}
\Sigma_{X}^{r}(E)=\left(H_{CX}-ES_{CX}\right)g_{XX}^{r}(E)\left(H_{XC}-ES_{XC}\right),\label{eq:Sigma}\end{equation}
 where \begin{equation}
g_{XX}^{r}(E)=\left[\left(E+i\eta\right)S_{XX}-H_{XX}\right]^{-1}\label{eq:grXX}\end{equation}
 is the retarded Green's function of region $X$. The matrix \begin{equation}
\Gamma_{X}(E)=-2\mathrm{Im}\left[\Sigma_{X}^{r}(E)\right]\label{eq:Gamma}\end{equation}
 is the line-broadening matrix. Note that both matrices $A_{X}(E)$
and $\Gamma_{X}(E)$ are positive-semidefinite. For notational convenience,
we will henceforth suppress the energy dependence of the quantities.

In the following we use a basis-independent notation with operators
such as $\hat{A}_{L}$ and $\hat{\Gamma}_{R}$, which are defined
by their matrix elements in the $C$ space. We also assume the existence
of a dual basis $|e^{j}\rangle$, satisfying $\langle e^{j}|e_{k}\rangle=\delta_{jk}$
and $\hat{1}=\sum_{j\in C}|e^{j}\rangle\langle e_{j}|$. Now let $(A_{L})^{jk}=\langle e^{j}|\hat{A}_{L}|e^{k}\rangle$
and $(\Gamma_{R})_{jk}=\langle e_{j}|\hat{\Gamma}_{R}|e_{k}\rangle$.
The matrix elements $(\Gamma_{R})_{jk}$ are the components of Eq.~(\ref{eq:Gamma}).
They are {}``covariant'', since the factors $H_{jk}-ES_{jk}$ in
Eq.~(\ref{eq:Sigma}) are covariant. The elements $(A_{L})^{jk}$
are also just the components of Eq.\ (\ref{eq:AX_def}). However,
they are {}``contravariant'', since the Green's functions $G_{CC}^{a}$
and $G_{CC}^{r}$ {[}see Eq.\ (\ref{eq:GrCC}){]} are defined as
the inverse of covariant matrices, i.e.\ $\sum_{k}[(E\pm i\eta)S_{jk}+H_{jk}]\langle e^{k}|\breve{G}^{r,a}|e^{l}\rangle=\delta_{jl}$.

Motivated by Eq.\ (\ref{eq:Transmission}) we also define the transmission
probability operator \begin{equation}
\hat{T}_{1}=\hat{A}_{L}\hat{\Gamma}_{R}.\label{eq:T1def}\end{equation}
 We will now show how the eigenchannel wavefunctions for waves coming
in from the left are conveniently obtained from the right eigenvectors
of $\hat{T}_{1}$ in the nonorthogonal basis by slightly reformulating
the procedure presented in Ref.\ \onlinecite{Paulsson2007}.

Consider the eigenvalue equation \begin{equation}
\hat{A}_{L}|\chi_{j}\rangle=\lambda_{j}|\chi_{j}\rangle.\label{eq:AL.ev1}\end{equation}
 The eigenvectors $|\chi_{j}\rangle$ of the Hermitian operator $\hat{A}_{L}$
are orthonormal ($\langle\chi_{j}|\chi_{k}\rangle=\delta_{jk}$).
Using them we define the states $|\tilde{\chi}_{j}\rangle=\sqrt{\frac{\lambda_{j}}{2\pi}}|\chi_{j}\rangle$
and the corresponding dual ones $|\tilde{\chi}^{j}\rangle=\sqrt{\frac{2\pi}{\lambda_{j}}}|\chi_{j}\rangle$
(for all $\lambda_{j}\neq0$) so that $\langle\tilde{\chi}^{j}|\tilde{\chi}_{k}\rangle=\delta_{jk}$.
It was shown in Ref.\ \onlinecite{Paulsson2007} that the states
$|\tilde{\chi}_{j}\rangle$ are the device part of orthogonal linear
combinations of energy-normalized scattering states, describing waves
coming in from the left lead. The transmission eigenchannels $|\phi_{n}\rangle$
can be expanded as \begin{equation}
|\phi_{n}\rangle=\sum_{j}|\tilde{\chi}_{j}\rangle c_{jn},\label{eq:phi_n_expand}\end{equation}
 with $c_{jn}=\langle\tilde{\chi}^{j}|\phi_{n}\rangle$. The coefficients
in this expansion may be found from the eigenvalue equation \begin{equation}
\hat{T}_{1}|\phi_{n}\rangle=\tau_{n}|\phi_{n}\rangle.\label{eq:T1eigval}\end{equation}
 Multiplying by $\langle\tilde{\chi}^{j}|$ on the left and using
Eqs.\ (\ref{eq:AL.ev1})-(\ref{eq:phi_n_expand}), this results in\cite{Paulsson2007}
\begin{equation}
2\pi\sum_{k}\langle\tilde{\chi}_{j}|\hat{\Gamma}_{R}|\tilde{\chi}_{k}\rangle c_{kn}=\tau_{n}c_{jn}.\label{eq:T1_mat}\end{equation}
 Employing the normalization condition $\sum_{k}c_{kj}^{*}c_{kl}=\delta_{jl}$,
the linear combination of Eq.~(\ref{eq:phi_n_expand}) preserves
the energy normalization of the left-incoming states. For $\langle\tilde{\chi}_{j}|\hat{\Gamma}_{R}|\tilde{\chi}_{m}\rangle=\sum_{k,l}\tilde{d}_{kj}^{*}(\Gamma_{R})_{kl}\tilde{d}_{lm}$
the coefficients $\tilde{d}_{lm}=\langle e^{l}|\tilde{\chi}_{m}\rangle=\sqrt{\frac{\lambda_{m}}{2\pi}}d_{lm}$
are determined from $\hat{A}_{L}$ by multiplying Eq.~(\ref{eq:AL.ev1})
by $\langle e_{j}|$ on the left and inserting $\hat{1}$. This leads
to the generalized eigenvalue problem \begin{equation}
\sum_{k,l,m}S_{jk}(A_{L})^{kl}S_{lm}d_{mn}=\lambda_{n}\sum_{k}S_{jk}d_{kn},\label{eq:ALeigval}\end{equation}
 with $d_{mn}=\langle e^{m}|\chi_{n}\rangle$ and $\sum_{k,l}d_{kj}^{*}S_{kl}d_{lm}=\delta_{jm}$.

Putting these results together, the explicit form of the eigenchannel
wavefunction for region $C$ in terms of the basis functions is obtained
from \begin{equation}
\langle\vec{r}|\phi_{n}\rangle=\sum_{j,k}\langle\vec{r}|e_{j}\rangle\tilde{d}_{jk}c_{kn}.\label{eq:eigchan_wf}\end{equation}
 Using Eqs.\ (\ref{eq:GrCC})-(\ref{eq:Gamma}) and (\ref{eq:T1_mat})-(\ref{eq:eigchan_wf})
the transmission eigenchannel wavefunction can be computed without
resorting to a L\"owdin transformation.

We note that the eigenvalues $\tau_{n}$ of $\hat{T}_{1}$ in Eq.~(\ref{eq:T1eigval})
are real, since $\hat{\Gamma}_{R}$ is a Hermitian operator {[}see
Eq.~(\ref{eq:T1_mat}){]}. It is also easy to show that they agree
with the eigenvalues of more symmetric, Hermitian transmission operators
of the form $\hat{T}_{2}=\hat{t}\hat{t}^{\dagger}$. Given $|\phi_{n}\rangle$
and $\tau_{n}$ from Eq.~(\ref{eq:T1eigval}) and assuming, e.g.,
$\hat{t}=\hat{\sqrt{\Gamma_{R}}}\hat{G}^{r}\hat{\sqrt{\Gamma_{L}}}$,
the states $|\phi'_{n}\rangle=\hat{\sqrt{\Gamma_{R}}}|\phi_{n}\rangle$
are eigenstates of $\hat{T}_{2}$ with the eigenvalues $\tau'_{n}=\tau_{n}$.
Furthermore, it is easy to prove\cite{Pauly2008} that the eigenvalues
satisfy $0\leq\tau_{n}\leq1$, as expected for transmission probabilities.

\section{Relation between the tight-binding and Lorentz model\label{sec:Rel-TBM-LM}}

The Lorentz model (LM) is frequently used to describe the transmission
in the field of molecular electronics. Typically, a Lorentzian function
is fitted to the resonance dominating the transmission at the Fermi
energy. Here, we discuss, in which situation the LM coincides with
the TBM.

We consider the non-Hermitian eigenvalue problem $(\hat{H}+\hat{\Sigma}^{r})|\mu\rangle=\lambda_{\mu}|\mu\rangle$
with $\lambda_{\mu}=\epsilon_{\mu}+i\gamma_{\mu}$, the symmetric
and Hermitian Hamilton operator $\hat{H}$, and the symmetric, but
non-Hermitian retarded self-energy operator $\hat{\Sigma}^{r}=\hat{\Sigma}_{L}^{r}+\hat{\Sigma}_{R}^{r}$
composed of contributions from the $L$ and $R$ electrodes. By $\langle\tilde{\mu}|$
we denote the left eigenstate with the same eigenvalue $\lambda_{\mu}$
as the corresponding right eigenstate $|\mu\rangle$, i.e.\ $\langle\tilde{\mu}|(\hat{H}+\hat{\Sigma}^{r})=\lambda_{\mu}\langle\tilde{\mu}|$.
The $C$ region is assumed to be identical to the molecule in the
TBM {[}see Fig.~\ref{fig:TE_TBM_LM}(a){]}. Using the spectral decomposition
of the Green's function in the expression for the energy-dependent
transmission $\tau(E)$ {[}see Eqs.~(\ref{eq:Transmission}) and
(\ref{eq:AX_def}){]}, we obtain \begin{equation}
\tau(E)=\sum_{\mu,\nu}\frac{\langle\tilde{\mu}|\hat{\Gamma}_{L}|\nu\rangle\langle\tilde{\nu}|\hat{\Gamma}_{R}|\mu\rangle}{(E-\epsilon_{\mu}-i\gamma_{\mu})(E-\epsilon_{\nu}+i\gamma_{\nu})},\label{eq:tauE_TBM}\end{equation}
where the sum is over all those eigenstates $|\mu\rangle$ of the
biphenyl which obtain a finite linewidth $\gamma_{\mu}\neq0$ by the
coupling to the electrodes and which hence contribute to the transport.

Let us now make the wide-band approximation and consider the energy-independent
expression $\hat{\Sigma}^{r}=-i(\hat{\Gamma}_{L}+\hat{\Gamma}_{R})/2$
to be a small perturbation. Within lowest-order perturbation theory
we obtain $\lambda_{\mu}=\epsilon_{\mu}^{0}+i\gamma_{\mu}$ with $\hat{H}|\mu^{0}\rangle=\epsilon_{\mu}^{0}|\mu^{0}\rangle$
and $\gamma_{\mu}=\langle\mu^{0}|\hat{\Sigma}^{r}|\mu^{0}\rangle$.
Additionally, we assume a symmetric coupling $(\Gamma_{L})_{\alpha\alpha}=(\Gamma_{R})_{\omega\omega}=\Gamma$,
where the indices $\alpha,\omega$ refer to those atoms of the biphenyl
backbone which are closest to the $L,R$ electrodes {[}see Fig.~\ref{fig:TE_TBM_LM}(a){]}
and where local basis states are understood to be orthogonal in the
spirit of the H\"uckel approximation. By exploiting the inversion
symmetry of the TBM, it follows that $\gamma_{\mu}=\langle\mu^{0}|\hat{\Gamma}_{X}|\mu^{0}\rangle$
since $\hat{M}^{2}=\hat{1}$, $\hat{M}\hat{H}\hat{M}=\hat{H}$, and
$\hat{M}\hat{\Gamma}_{L}\hat{M}=\hat{\Gamma}_{R}$ with the operator
$\hat{M}=\hat{M}^{\dagger}$ describing the inversion of the molecule.

The perturbation theory is valid in the regime $\Gamma\ll t$, where
$t$ determines the separation between the resonance energies $\epsilon_{\mu}$
relevant for transport. When they are well separated, the largest
contributions to the transmission in Eq.~(\ref{eq:tauE_TBM}) arise
when $\mu=\nu$, since cross-terms are suppressed by a large off-resonant
denominator. In this case the transmission is well represented as
the sum of incoherent Lorentz resonances \begin{equation}
\tau(E)\approx\sum_{\mu}\frac{\gamma_{\mu}^{2}}{(E-\epsilon_{\mu})^{2}+\gamma_{\mu}^{2}},\label{eq:TE_LM}\end{equation}
and the TBM simplifies to the LM.

\bibliographystyle{apsrev}

\end{document}